\documentclass[twocolumn,showpacs,prb]{revtex4}

\usepackage{epsfig,graphicx,hhline}

\usepackage{dcolumn}% Align table columns on decimal point
\usepackage{bm}% bold math\\

\begin{document}

\title{First-principles study of cubic perovskites SrMO$_3$ (M = Ti, V, Zr and Nb)}

\author{I.R. Shein$^*$, V.L.Kozhevnikov and A.L. Ivanovskii}

\affiliation {Institute of Solid State Chemistry, Ural Branch of
the Russian Academy of Sciences, 620219, Ekaterinburg, Russia}

\begin{abstract}
Using the full-potential linearized-augmented-plane-wave (FLAPW)
method, we have analyzed systematically the trends in the
structural and electronic properties of the 3d and 4d
transition-metal oxides SrMO$_3$ (M = Ti, V, Zr and Nb). The
optimized lattice parameters, bulk modules, densities of states,
band structures and charge density distributions are obtained and
compared with the available theoretical and experimental data.
The energy gap between O2p - Md bands increases as the covalency
of the system decreases going from 3d to 4d based perovskites.
The electron configurations of Sr(Ti,Zr)O$_3$ and Sr(V,Nb)O$_3$
usually referred to as d$^0$ and d$^1$ oxides, respectively,
differ considerably from these idealized "ionic" configurations,
and the deviations increase with increasing of the d-p covalent
overlap in the oxides.

$^*$ E-mail: shein@ihim.uran.ru
\end{abstract}


% 74.70.-b Superconducting materials (excluding high-Tc compounds)
% 71.20.Dg Alkali and alkaline earth metals
% 71.20.-b Electron density of states and band structure of
%          crystalline solids

\maketitle I. INTRODUCTION\\

Perovskites SrMO$_3$ attract much attention from physicists and
material scientists because of the unusual combination of their
magnetic, electronic and transport
properties\cite{Leonidov,Hong,Takeno,Yaremchenko,Kozhevnikov,Shaula}.
However, the electronic structure features in these complicated
systems are not understood comprehensively yet, and attempts to
explain their behavior in different conditions are often based on
oversimplified models. Through quite a number of publications are
known on strontium oxide perovskites with 3d metals on M
sites\cite{Arai,Marques,Bottin,Guo,Mete,Matar1,Matar2,Takahashi,Mattheiss,Pertosa,Perkins,Krasovska,Jarlborg},
the research focus has long been on strontium titanate, SrTiO$_3$
- one of the generic representatives of 3d transition-metal oxide
perovskites. For instance, the LCAO-based approaches are used in
non-self-consistent or self-consistent
calculations\cite{Mattheiss,Pertosa,Perkins}. Recently more
accurate self-consistent full-potential LAPW, LMTO, ab-initio
pseudopotential band structure calculations are
performed\cite{Arai,Marques,Bottin,Guo,Matar1,Krasovska,Jarlborg}.
Some attention is given to the related materials, viz., the 3d$^1$
metal-like perovskite SrVO$_3$, solid solutions
SrTi$_{1-x}$V$_x$O$_3$, SrVO$_3$/SrTiO$_3$ superlattices and
heteroepitaxial structures SrVO$_3$/SrTiO$_3$/Si\cite{Moon,Kim}.
Also, the cubic SrVO$_3$ is treated as a model system for
discussion of orbital fluctuations\cite{Pavarini} and correlation
effects\cite{Nekrasov}. As an example of 4d based perovskites,
the cubic SrZrO$_3$ is studied by the discrete variation (DV)
molecular orbital method\cite{Yoshino} and, recently the band
structure calculations are performed\cite{Mete} within the
local-density approximation (LDA).\\
Finally, the electronic properties of the ideal cubic SrNbO$_3$,
which may be considered as a prototype of an ample family of
perovskite-related layered niobates\cite{Lichtenberg}, are
investigated in work\cite{Turzhevsky} by means of LMTO-ASA
approach. In this Communication, we report results of a
systematic study of electronic properties of perovskite-type 3d
and 4d transition-metal oxides SrMO$_3$ (M = Ti, V, Zr and Nb) by
means of the FLAPW method within the generalized gradient
approximation (GGA). We have evaluated a number of physical
parameters such as optimized crystal structure, bulk modulus,
density of states, band structure and electron density
distribution and considered their variations with M-type metal.\\

II. MODELS AND METHOD\\

The considered SrMO$_3$ perovskites are assumed to have ideal
cubic structure (s.g. Pm3m) where atomic positions in the
elementary cell are M: 1a (0,0,0); O: 3d (0,0,1/2); and Sr: 1b
(1/2,1/2,1/2). The electronic configurations are taken
Ar3d$^2$4s$^2$ for Ti, Ar3d$^3$4s$^2$ for V, Kr4d$^2$5s$^2$ for
Zr, Kr4d$^3$5s$^2$ for Nb, Kr5s$^2$ for Sr and He2s$^2$2p$^4$ for
O. Here, the noble gas cores are distinguished from the
sub-shells of valence electrons. The equilibrium structural
parameters and bulk modules are calculated using the Vienna
package WIEN2k, which is an implementation of the hybrid full
potential (linear) augmented plane-wave plus local orbitals
(L)APW+lo method within the density-functional
theory\cite{Blaha}. The basis set inside each MT sphere is split
into core and valence subsets. The core states are treated within
the spherical part of the potential only, and are assumed to have
a spherically symmetric charge density confined within MT
spheres. The valence part is treated with the potential expanded
into spherical harmonics up to l = 4. The valence wave functions
inside the spheres are expanded up to l = 12. The plane-wave
expansion with RMT*KMAX equal to 10 and k sampling with 10*10*10
k-points mesh in the Brillouin zone (BZ) is used. All
calculations are carried out with optimized lattice constants. We
use the Perdew-Burke-Ernzerhof generalized gradient approximation
(GGA)\cite{Perdew96} for the exchange correlation potentials.
Relativistic effects are taken into account within the
scalar-relativistic approximation.\\
The self-consistent calculations were considered to have
converged when the difference in the total energy of the crystal
did not exceed 0.1 mRy as calculated at consecutive steps. We
have adopted the MT radii of 1.7 a.u. for O, 1.8 a.u. for M, and
2.5 a.u. for Sr. The lattice constants and bulk modules are
calculated by fitting the total energy versus volume according to
the Murnaghan's equation of state\cite{Murnaghan}. The total
density of states (DOS) was obtained using a modified tetrahedron
method\cite{Blochl}.\\

III. RESULTS AND DISCUSSION\\

The lattice constants of the selected perovskites SrMO$_3$, where
M = Ti, Zr, V and Nb, were calculated under constrain of cubic
symmetry. The results and some available experimental and
theoretical data are summarized in Table 1. It is seen that for
SrTiO$_3$ and SrZrO$_3$ the calculated LDA constants\cite{Mete}
are smaller compared to experimental ones. The GGA improves the
lattice constants over the LDA. In fact, the obtained GGA lattice
constants are very close to experimental data; the deviations for
SrTiO$_3$, SrZrO$_3$ and SrVO$_3$ achieve only about 1\%. The
bulk modules decrease going from SrTiO$_3$ to SrZrO$_3$ and from
SrVO$_3$ to SrNbO$_3$; and B$_0$ values for metal-like SrVO$_3$
and SrNbO$_3$ are higher than for insulating SrTiO$_3$ and
SrZrO$_3$ perovskites. These trends correlate with changes in the
lattice constants.\\

\begin{table}
\begin{center}
\caption{ Calculated and experimental values for the lattice
parameters, bulk modules and its pressure derivative of SrMO$_3$
(M= Ti, V, Zr and Nb).}
\
\
\begin{tabular}{|c|c|c|c|c|c|}
\hline
& \multicolumn{2}{c|}{ }&\multicolumn{2}{c|}{ }&{ }\\
&\multicolumn{2}{c|}{Lattice}&\multicolumn{2}{c|}{Bulk  modulus}&{B$_0$'}\\
Parameters &\multicolumn{2}{c|}{lattice (\AA)}&\multicolumn{2}{c|}{ B$_0$ (GPA)}&{ }\\
\cline{2-5} & & & & & \\ & Calcul. &Exper.& Calcul.& Exper. & { }\\
& & & & & \\
\hline 
& & & & & \\
SrTiO$_3$ & 3.9456*& 3.905\cite{Data}& 167.9* & 183\cite{Data}& 4.56*\\
            & 3.878\cite{Mete}&            & 191\cite{Mete}&        &      \\
& & & & & \\    
\hline 
& & & & & \\
SrVO$_3$ & 3.8662* & 3.841\cite{Rey} & 182.8*& - & 5.11*\\
                &         &3.8425\cite{Lan} &       &   & \\
& & & & & \\
\hline 
& & & & & \\
SrZrO$_3$ &4.1794*&4.109\cite{Smith}&158.28*&150\cite{Ligny}&3.66*\\
                 &4.095\cite{Mete}&  &171\cite{Mete}& & \\
& & & & & \\
\hline 
& & & & & \\
SrNbO$_3$ & 4.0730* & - & 170.0* & - & 3.71*\\
& & & & & \\
\hline
\end{tabular}
\end{center}
\begin{flushleft}
* - our FLAPW-GGA data\\
\end{flushleft}
\end{table}

The dispersion E(k) in the high-symmetry directions in the
Brillouin zone are shown in Fig.1 where similarity in the energy
bands is evident in the oxide perovskites under study. The general
features in the spectra can be specified in more detail with the
using of SrTiO3 as example. The lower bands, which are not shown
in Fig.1, contain 12 electrons in Sr4p and O2s semi-core states
located about -15.5 eV  below the Fermi level. The upper valence
band (VB) with the width of about 4.6 eV is derived mainly from
O2p orbitals with some admixture of Ti and Sr states, and it
contains 18 electrons. The bottom of the conduction band (CB) is
composed mainly of Ti3d t$_{2g}$ and e$_g$ states, and at higher
energies the Sr4d - like bands are placed. The valence-band top of
SrTiO$_3$ is composed of O2p orbitals, and it is located at the R
point, while the VB maximum at the M point is about 0.09 eV lower
than the R. The conduction-band bottom is at the $\Gamma$ point.
The CB minimum at X is by 0.11 eV higher than at the $\Gamma$
point. Thus the indirect gap between R and $\Gamma$ points is 1.81
eV, and the direct gap at $\Gamma$ is 2.16 eV. These features of
the band structure agree well with the previous calculations, see
Table 2. At the same time the experimentally determined indirect
band gap in SrTiO$_3$ achieves about 3.30 eV, while the direct
band gap energy is equal to 3.75 eV\cite{Benthem,Lee}. The
divergence is related to the well-known underestimation of the
band gap values within LDA - based calculation methods that treat
electron exchange correlations in an approximate way. Notice,
however, that the difference in the calculated values for the
direct and indirect gaps equals exactly to the same difference in
the experimentally observed band gaps.\\
The nine occupied bands at the $\Gamma$ point consist of three
triply degenerate levels ($\Gamma_{15}$, $\Gamma_{25}$ and
$\Gamma_{15}$) with energies of -2.82, -1.17 and -0.39 eV below
E$_F$, respectively. The splitting of the levels ($\Gamma_{15}$ -
$\Gamma_{25}$) and ($\Gamma_{25}$ - $\Gamma_{15}$) is due to the
crystal field effects and electrostatic interaction between O2p
orbitals. The triply ($\Gamma_{25}'$, +1.81 eV) and the doubly
($\Gamma_{12}$, +4.12 eV) degenerate levels in the conduction
band originate from t$_{2g}$ ($\pi$*) and e$_g$ ($\sigma$*) bands
of titanium, and they are separated with the energy gap of 2.31 eV.\\
The calculated total (TDOS) and partial (PDOS) densities of states
are shown in Fig.2 within the energy interval [E$_F$ - 6 eV $\div$
E$_F$ + 8 eV]. The TDOS is almost entirely composed from the PDOSs
for Ti3d and O2p states. The valence TDOS originates predominantly
from O2p states, which is consistent with SrTiO$_3$ being a formal
d$^0$ system with titanium in 4+ oxidation state. There is also a
small contribution in the valence TDOS from strontium 5s states.
However, there is a Ti3d PDOS in the occupied energy range due to
the hybridization with the O2p states indicating the presence of a
covalent bonding. Note also that within the cubic symmetry of
SrTiO$_3$ the hybridization between t$_{2g}$ and e$_g$ states is
forbidden, and the orbitals within both bands are degenerate.
Nevertheless, because of the Ti-O hybridization the admixture of
the both t$_2g$ and e$_g$ states to the oxygen 2p bands and also
their partial overlapping occur. It is seen from Fig. 2 that the
eg states (directed towards oxygen atoms so that they form
$\sigma$ bonds with respective O 2p orbitals) have lower energy in
comparison with Ti t2g states forming $\pi$-like bonds with O 2p states.\\

\begin{table}
\begin{center}
\caption{Calculated and experimental values for the bandwidths (in eV) of SrMO$_3$ (M= Ti, V, Zr and Nb).}
\
\
\begin{tabular}{|c|c|c|c|c|}
\hline
 & & & & \\
Parameters & SrTiO $_3$ & SrZrO$_3$ & SrVO$_3$ & SrNbO$_3$\\
 & & & & \\
\hline 
 & & & & \\
Valence band & 4.59*& 3.86* & 7.18* & 8.12*\\
& 4.95\cite{Mete}&4.32\cite{Mete}&7.5\cite{Nekrasov}&6.12\cite{Turzhevsky}\\
 & & & & \\
\hline 
 & & & & \\
Conduction band&-&-&1.05*&1.21*\\
(occupied part)&-&-&1.1\cite{Nekrasov}&1.18\cite{Turzhevsky}\\
 & & & & \\
\hline
& & & & \\ 
 Band gap:& & & &\\
direct($\Gamma$) &2.16*&3.75*& 1.40*& 2.79*\\
  &2.30\cite{Mete}&3.62\cite{Mete}& &\\
  &3.75\cite{Benthem}& & &\\
indirect(R-$\Gamma$)& 1.81*& 3.30* & 1.02* & 2.38\\  
  & 1.92\cite{Mete}&3.37\cite{Mete}& & 2.31\cite{Turzhevsky}\\
  & 3.25\cite{Benthem} & 5.6\cite{Lee} & &\\
  &3.4\cite{Lee}& & &\\
& & & & \\
\hline
\end{tabular}
\end{center}
\begin{flushleft}
* - our FLAPW-GGA data\\
** - Experimental data\cite{Benthem}\cite{Lee}
\end{flushleft}
\end{table}

The TDOS profile is formed basically of four peaks centered at -
3.1, -1.3, + 3.6 and + 5.1 eV. A very flat O2p - like band (along
$\Gamma$-X-M-$\Gamma$, see Fig. 1) is responsible for the highest
of them placed at - 3.1 eV. The contribution of Ti3d states is
quite small in the peak near -1.3 eV. Therefore the DOS peak at
this energy may be attributed to quasi-flat bands originating
primarily from non-bonding O2p states. The peaks in CB centered
at + 3.6 and + 5.1 eV are formed by significant contributions
from t$_{2g}$ and e$_g$ bands of Ti3d states.\\
The comparison of the band structures of the isoelectronic and
isostructural SrTiO$_3$ and SrZrO$_3$ shows that the dispersion
of the bands in SrTiO$_3$ is much stronger than in SrZrO$_3$. For
example, the VB bandwidth in SrTiO$_3$ is about 0.73 eV larger.
This means that Ti-O bonds are more covalent than respective
bonds between zirconium and oxygen. Another obvious difference is
that the band gap (which remains indirect in SrZrO$_3$; the
bottom of the conduction band is located at the $\Gamma$ point in
the cubic BZ while the top of the valence band is at the R point)
is about 1.49 eV (45\%) larger than that of SrTiO$_3$. In the case
of SrZrO$_3$ the splitting of the bands ($\Gamma_{15}$ -
$\Gamma_{25}$) and ($\Gamma_{25}$ - $\Gamma_{15}$) is about 0.75
and 0.62 eV, respectively.\\
Band structures for d$^1$ perovskites SrVO$_3$ and SrNbO$_3$ are
presented in Fig.1, and band structure parameters are listed in
Table 2. The VBs of SrVO$_3$ and SrNbO$_3$ consist of completely
occupied oxygen 2p bands and partially occupied (V,Nb)3d bands.
The electronic states near the Fermi level originate mainly from
3d (t$_{2g}$) states.\\
One can see from Fig.2 that for SrVO$_3$ and SrNbO$_3$ there is
some contribution of the t$_{2g}$ and e$_g$ bands in the energy
intervals from -7.2 to -2.1 eV  and from -8.3 to  3.6 eV,
respectively, because of the hybridization with O2p states. In
the case of SrVO$_3$ the splitting of the bands ($\Gamma_{15}$ -
$\Gamma_{25}$) and ($\Gamma_{25}$ - $\Gamma_{15}$) are about 1.94
and 0.81 eV; for SrNbO$_3$ these values are about 0.95 and 0.74
eV, i.e. larger than for SrTiO$_3$ and SrZrO$_3$, respectively.
For SrVO$_3$ and SrNbO$_3$ the t$_{2g}$ states are dominant in the
vicinity of the Fermi level. The e$_g$ and t$_{2g}$ bands are well
separated in the CB, and the maximum of the DOS in the eg band is
centered above the upper band edge of the t$_{2g}$ band. The main
weight of t$_{2g}$ and e$_g$ states in SrVO$_3$ is observed in the
energy intervals [ 1.1 $\div$ +1.41 eV] and [+1.38 $\div$ +5.8
eV], respectively, while in SrNbO3 the respective regions are
located at [-1.2 $\div$ +2.3 eV] and [+1.61 $\div$ + 8.5 eV].\\
The obtained results show that the total density of states at the
Fermi level N$_{tot}$(E$_F$) in SrVO$_3$ (1.678 states/eV/cell)
is about 20\% higher than in SrNbO$_3$ (1.256 states/eV/cell).
The main contribution to N$_{tot}$(E$_F$) in both systems
originates from the t$_2g$ bands, and the N$_{tot}$(E$_F$) changes
are also controlled by the contributions from t$_{2g}$ states:
going from SrVO$_3$ to SrNbO$_3$, these contributions decrease
rapidly from 1.109 to 0.542 states/eV/cell. In order to
illustrate the bonding picture in systems under consideration we
have plotted the charge density maps for SrTiO$_3$ and SrVO$_3$,
Fig. 3. It is seen that the covalent interaction between Ti and O
or V and O atoms is very strong.\\

\begin{table}
\begin{center}
\caption{Occupation indices (in e) of M e$_g$ and t$_{2g}$ bands of SrMO$_3$ (M= Ti, V, Zr and Nb).}
\
\
\begin{tabular}{|c|c|c|c|c|}
\hline
 & & & & \\
Band/system & SrTiO $_3$ & SrZrO$_3$ & SrVO$_3$ & SrNbO$_3$\\
 & & & & \\
\hline
 & & & & \\
 M e$_g$ & 0.642*& 0.742 & 0.307 & 0.533\\
M t$_{2g}$ &0.473 & 1.313 & 0.234 & 0.791 \\
 & & & & \\
\hline
\end{tabular}
\end{center}
\end{table}

Finally, the perovskites SrTiO$_3$ and SrVO$_3$ (and their 4d
analogues) are usually referred to as d$^0$ and d$^1$ oxides in
the literature. It is seen, however, from Table 3 that calculated
electron configurations (viz., Ti3d$^{1.115}$, Zr4d$^{0.541}$,
V3d$^{2.055}$ and Nb4d$^{1.324}$) differ considerably from the
idealized "ionic" configurations, and the respective deviations
increase with the d-p covalent overlap in the oxides.\\

IV. CONCLUSIONS\\

In summary, we have performed FLAPW calculations in order to
systematically study ground state properties and electronic
structure features in 3d and 4d transition-metal oxides SrMO$_3$,
where M = Ti, V, Zr and Nb. The energy gap between 2p oxygen and
d metal bands increases with the decrease in the covalent p-d
overlap. The optimized lattice parameters, bulk modules,
densities of states, band structures and charge density
distributions are obtained and compared with the available
theoretical and experimental data. The analysis of the d band
population reveals that the deviation of the d metal electron
configurations from the formal ionic state (d$^0$ for SrTiO$_3$,
SrZrO$_3$ and d$^1$ for SrVO$_3$ and SrNbO$_3$) increases with p-d
covalent overlap.\\

ACKNOWLEDGEMENTS\\

The work was supported by Russian Academy of Science within the
program "Hydrogen energy and fuel elements".

%%%%%%%%%%%%%%%%%%%%%%%%%%%% Bibliography %%%%%%%%%%%%%%%%%%%%%%%%%%%

\begin{figure*}[!htb]
\vskip  0cm
\begin{tabular}{c}

\includegraphics[width=15.0 cm,clip]{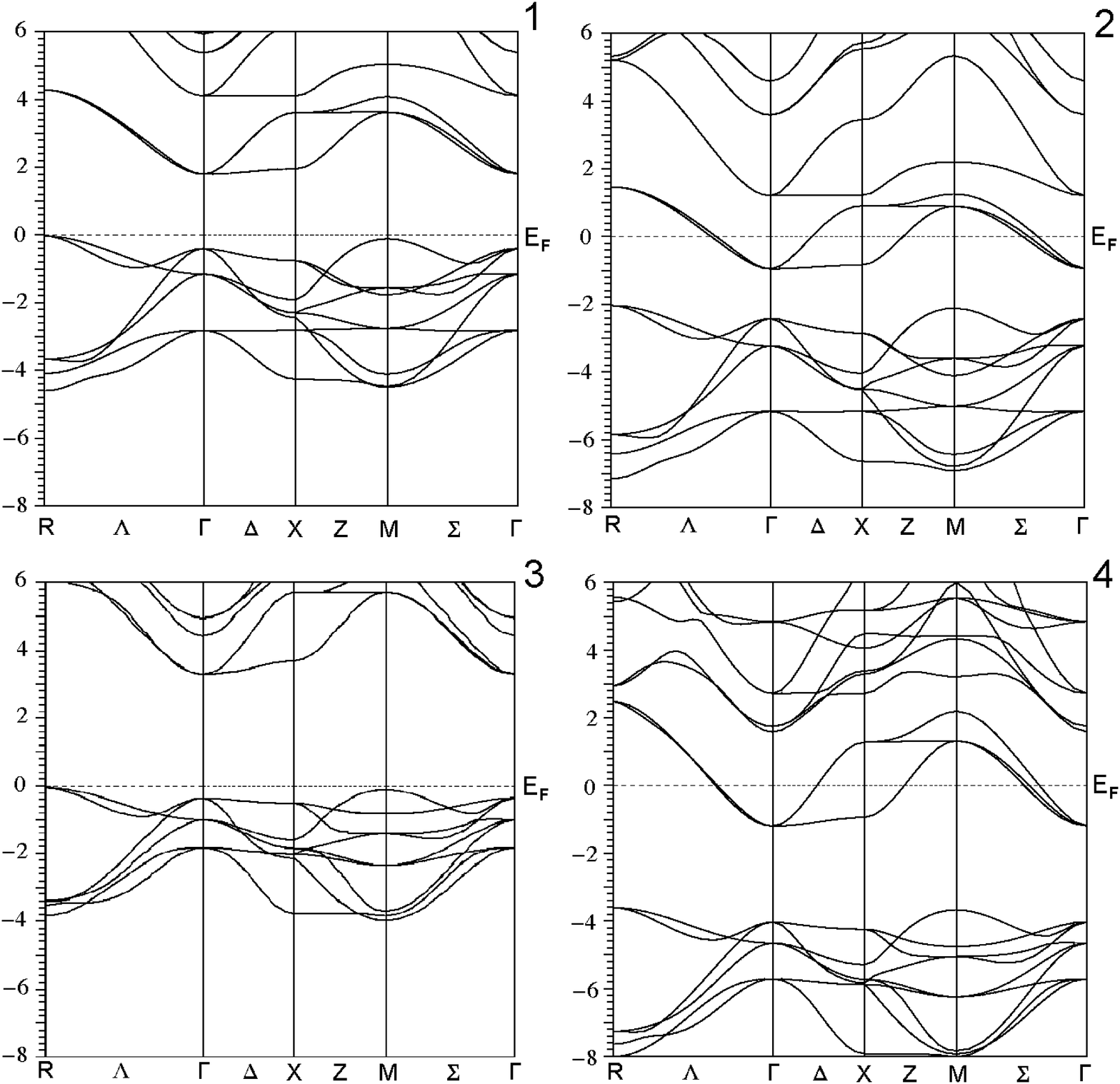}\\

\end{tabular}

\vspace{-0.02cm} \caption[a] { \small Band structures of cubic perovskites: 1- SrTiO$_3$ 2- SrVO$_3$ 3- SrZrO$_3$, 4- SrNbO$_3$. The Fermi level corresponds to 0.0 eV.}
%\end{floatingfigure}
\end{figure*}

\begin{figure*}[!htb]
\vskip  0cm
\begin{tabular}{c}

\includegraphics[width=12.0 cm,clip]{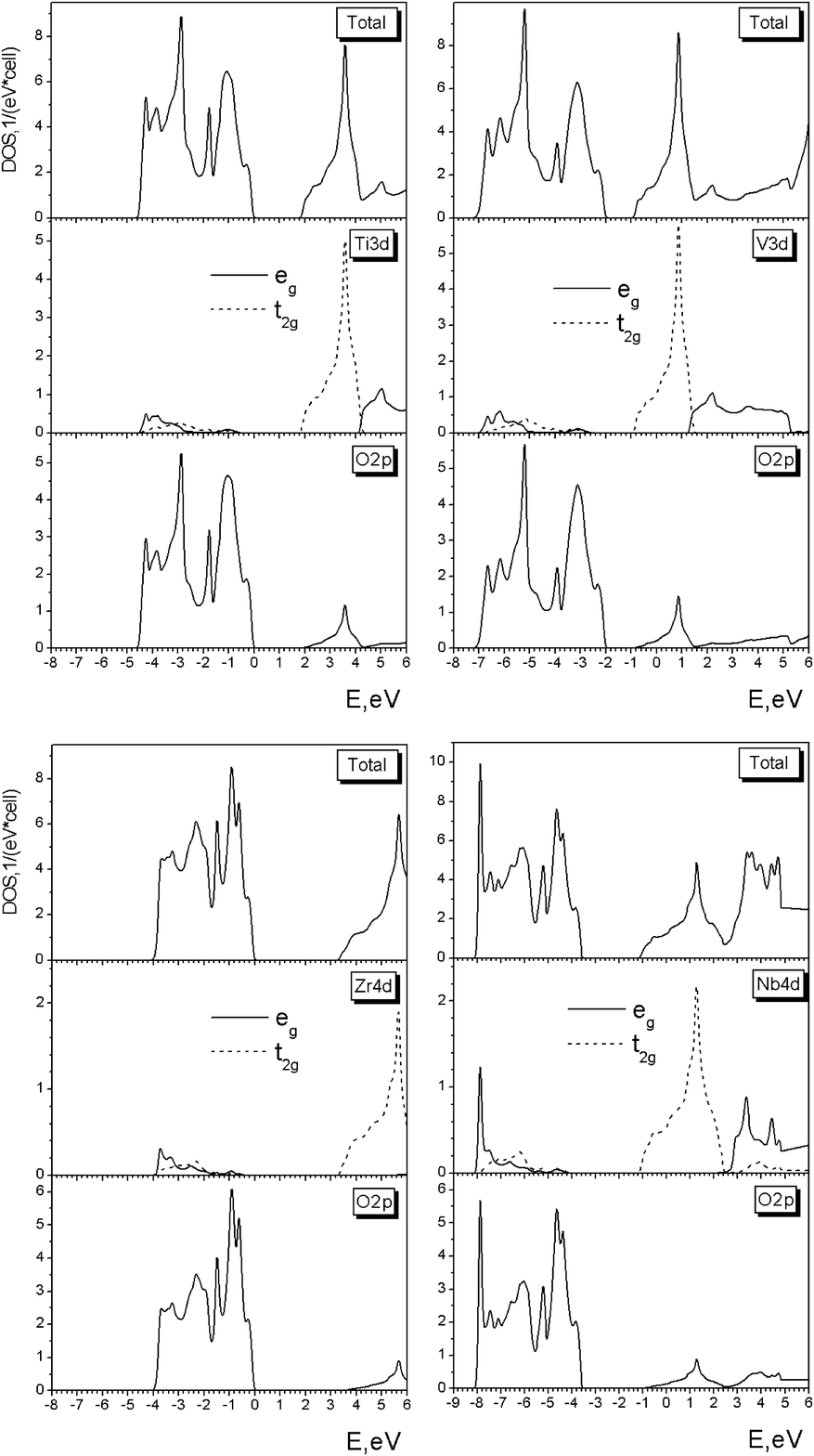}\\

\end{tabular}

\vspace{-0.02cm} \caption[a] { \small Total and projected DOSs for the cubic perovskites SrBO$_3$ (B = Ti, V, Zr and Nb).}
%\end{floatingfigure}
\end{figure*}

\begin{figure*}[!htb]
\vskip  0cm
\begin{tabular}{c}

\includegraphics[width=10.0 cm,clip]{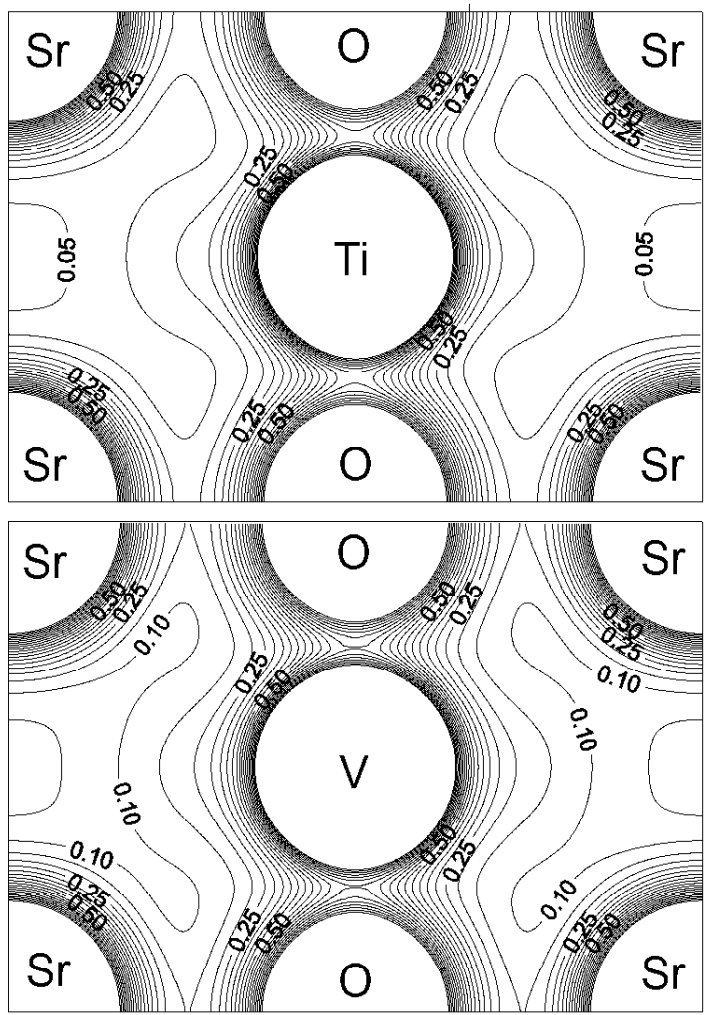}\\

\end{tabular}

\vspace{-0.02cm} \caption[a] { \small Charge densities maps for SrTiO$_3$ and SrVO$_3$ in (100) plane.}
%\end{floatingfigure}
\end{figure*}

\end{document}